\documentclass[conference]{IEEEtran}
\IEEEoverridecommandlockouts
\usepackage{cite}
\usepackage{amsmath,amssymb,amsfonts}
\title{Aligned mathematics}
\usepackage{algorithm} 
\usepackage{algpseudocode} 
\usepackage{graphicx}
\usepackage{textcomp}
\usepackage{xcolor}

\def\BibTeX{{\rm B\kern-.05em{\sc i\kern-.025em b}\kern-.08em
    T\kern-.1667em\lower.7ex\hbox{E}\kern-.125emX}}

\begin{document}

\title{Active Inference for Sum Rate Maximization in UAV-Assisted Cognitive NOMA Networks}
\author{felix.obite }
\date{August 2022}

\author{\IEEEauthorblockN{Felix~Obite\textsuperscript{1,2}, Ali~Krayani\textsuperscript{1}, Atm~S.~Alam\textsuperscript{2}, Lucio~Marcenaro\textsuperscript{1}, Arumugam~Nallanathan\textsuperscript{2}, Carlo~Regazzoni\textsuperscript{1}}
\IEEEauthorblockA{\textsuperscript{1}\text{DITEN, University of Genova, Italy} \\
\textsuperscript {2}\text{EECS, Queen Mary University of London, United Kingdom}\\
\small emails:felix.obite@edu.unige.it, ali.krayani@ieee.org, \{lucio.marcenaro, carlo.regazzoni\}@unige.it,\                                     \{a.alam, a.nallanathan\}@qmul.ac.uk }
}

\maketitle

\begin{abstract}
Given the surge in wireless data traffic driven by the emerging Internet of Things (IoT), unmanned aerial vehicles (UAVs), cognitive radio (CR), and non-orthogonal multiple access (NOMA) have been recognized as promising techniques to overcome massive connectivity issues. As a result, there is an increasing need to intelligently improve the channel capacity of future wireless networks. Motivated by active inference from cognitive neuroscience, this paper investigates joint subchannel and power allocation for an uplink UAV-assisted cognitive NOMA network. Maximizing the sum rate is often a highly challenging optimization problem due to dynamic network conditions and power constraints. 
To address this challenge, we propose an active inference-based algorithm. We transform the sum rate maximization problem into abnormality minimization by utilizing a generalized state-space model to characterize the time-changing network environment. The problem is then solved using an Active Generalized Dynamic Bayesian Network (Active-GDBN). The proposed framework consists of an offline perception stage, in which a UAV employs a hierarchical GDBN structure to learn an optimal generative model of discrete subchannels and continuous power allocation. In the online active inference stage, the UAV dynamically selects discrete subchannels and continuous power to maximize the sum rate of secondary users. By leveraging the errors in each episode, the UAV can adapt its resource allocation policies and belief updating to improve its performance over time. Simulation results demonstrate the effectiveness of our proposed algorithm in terms of cumulative sum rate compared to benchmark schemes.
\end{abstract}


\section{Introduction}
The present and emerging wireless technologies, such as 6G, are expected to experience an increase in data intensity. There is a growing expectation to connect a larger number of self-autonomous devices and Internet of Things (IoT) devices, putting pressure on the existing wireless network  \cite{bonati2020open}. To address these demands, future wireless systems need to incorporate innovative technologies like unmanned aerial vehicles (UAVs), cognitive radio (CR), and non-orthogonal multiple access (NOMA). The integration of these technologies requires intelligent resource allocation to improve system performance. UAVs have gained significant attention in wireless communications due to their advantageous line-of-sight (LoS) communication, cost-effectiveness, miniaturization, and flexibility
\cite{luong2021deep}. CR serves as the main technology that enables secondary users (SUs) to utilize licensed spectrum when available without causing interference to primary users (PUs) \cite{shen2019uav}. Conversely, there has been a shift towards NOMA as an alternative to orthogonal multiple access (OMA) to overcome its limitations. NOMA has demonstrated superior spectrum efficiency, user fairness, and the ability to accommodate multiple users simultaneously in the same sub-channel by employing superposition coding (SC) at the transmitter and successive interference cancellation (SIC) at the receiver \cite{sohail2018non}. 

However, in order to fully harness the promised benefits of NOMA, a key challenge lies in jointly optimizing discrete subchannels and continuous power to maximize the sum rate in such a dynamic system. Additionally, it has been established that the problem of maximizing the sum rate in wireless networks is strongly NP-hard \cite{salaun2018optimal}. Hence, numerous suboptimal or heuristic approaches have been proposed by researchers. The authors in \cite{chen2015suboptimal} explore heuristic and iterative search optimization methods for user pairing and resource allocation in the uplink NOMA scenario. In \cite{guo2020weighted}, the authors investigate the stochastic successive convex approximation method to maximize the sum rate of users under imperfect channel state conditions.
In \cite{lei2016power}, an upper bound is derived for the optimal weighted sum rate, and the authors propose a near-optimal approach using Lagrangian duality and dynamic programming. For a comprehensive review of conventional optimization approaches, we refer the readers to \cite{he2019joint}. It is important to note that these traditional optimization schemes lack adaptive online self-awareness and often involve complex mathematical formulations, making them impractical for real-time systems that require minimal latency.

In recent years, machine learning techniques have demonstrated significant potential in addressing complex computational tasks and have been widely implemented in various wireless communication systems\cite{you2019ai}. However, deep learning methods require well-labeled datasets for training in order to achieve accurate results. Obtaining such datasets can be challenging in complex wireless networks, and the resulting models may be difficult to interpret \cite{kohoutova2020toward}. Likewise, in \cite{he2019joint}, deep reinforcement learning (Deep RL) is employed to maximize the sum rate, power allocation, and channel assignment in a multi-carrier NOMA scheme. Nevertheless, despite the recent successes of RL, several limitations hinder its full implementation in dynamic systems \cite{irpan2018deep}. Firstly, RL algorithms often require numerous iterations to converge to an optimal solution due to the strong influence of negative rewards
\cite{friston2009reinforcement, kurenkov2018reinforcement}. An RL agent must take several bad actions in order to learn how to improve its policy. Additionally, RL agents are typically trained for specific predefined tasks, which limits their ability to generalize to new experiences. Generalizing to new experiences necessitates retraining or modifying the agent, or incorporating meta-learning capabilities \cite{hospedales2021meta}.
An alternative approach widely studied and rooted in neurocognitive science, called active inference, provides a fundamental framework for characterizing adaptive behaviors in unknown and complex environments \cite{friston2009reinforcement, 9829873}. 
In this framework, every agent (considered a self-organizing system) maintains a dynamic equilibrium with its external environment to minimize prediction errors
\cite{friston2006free}. Preliminary results suggest that active inference is more adaptable and resilient in a variety of settings that are challenging for RL models \cite{tschantz2020reinforcement}.

We also observe that the majority of active inference agents are trained to learn generative models with predefined sections of the state space \cite{friston2009reinforcement, sajid2022active}. While this approach is suitable for discrete state spaces, it becomes impractical for complex dynamic systems
 \cite{ccatal2020learning}. In this paper, we explore active inference using a unique generalized dynamic Bayesian network (Active-GDBN) to learn the complex, time-changing network environment. The key contributions of this study are summarized as follows:
\begin{itemize}
    \item We have developed and implemented an active inference-based algorithm called Active-GDBN to address the sum rate maximization problem in a UAV-assisted cognitive NOMA network. In this algorithm, the UAV is equipped with a generative model that is learned offline, capturing the dynamic rules that generate preferred observations (i.e., optimal superimposed signals). This learned knowledge serves as a prior target when the UAV becomes active during the online deployment process. We describe how the UAV dynamically learns both discrete subchannels and a continuous power allocation policy online to minimize prediction errors or abnormalities.
    \item We formulate the problem of maximizing the sum rate as a challenge of minimizing abnormalities, employing a generalized state-space formulation to capture the temporal dynamics of the radio environment. Unlike most existing papers, which discretize power allocation, our proposed framework optimizes continuous power. Discretizing power allocation introduces quantization errors and increases computational complexity \cite{wang2020drl}. Additionally, our algorithm is explainable because it estimates and represents the dynamic causal structure of the training environment at both discrete and continuous states. 
    \item The numerical findings using simulated data provide evidence of the efficiency of our proposed algorithm in achieving a higher cumulative sum rate compared to benchmark schemes.
    \end{itemize}

The remainder of this paper is organized as follows: Section \ref{Sec_systemModel_probForm} describes the system model and problem formulation. The proposed method for joint sub-channel and power allocation is described in Section \ref{Sec_proposedMethod}. The simulation results and discussion are presented in Section \ref{Sec_simulationResults}. Section \ref{Sec_conclusion} concludes the paper.

\section{System Model and Problem Formulation} \label{Sec_systemModel_probForm}
\begin{figure}[htp]
    \centering
    \includegraphics[width=7.0cm]{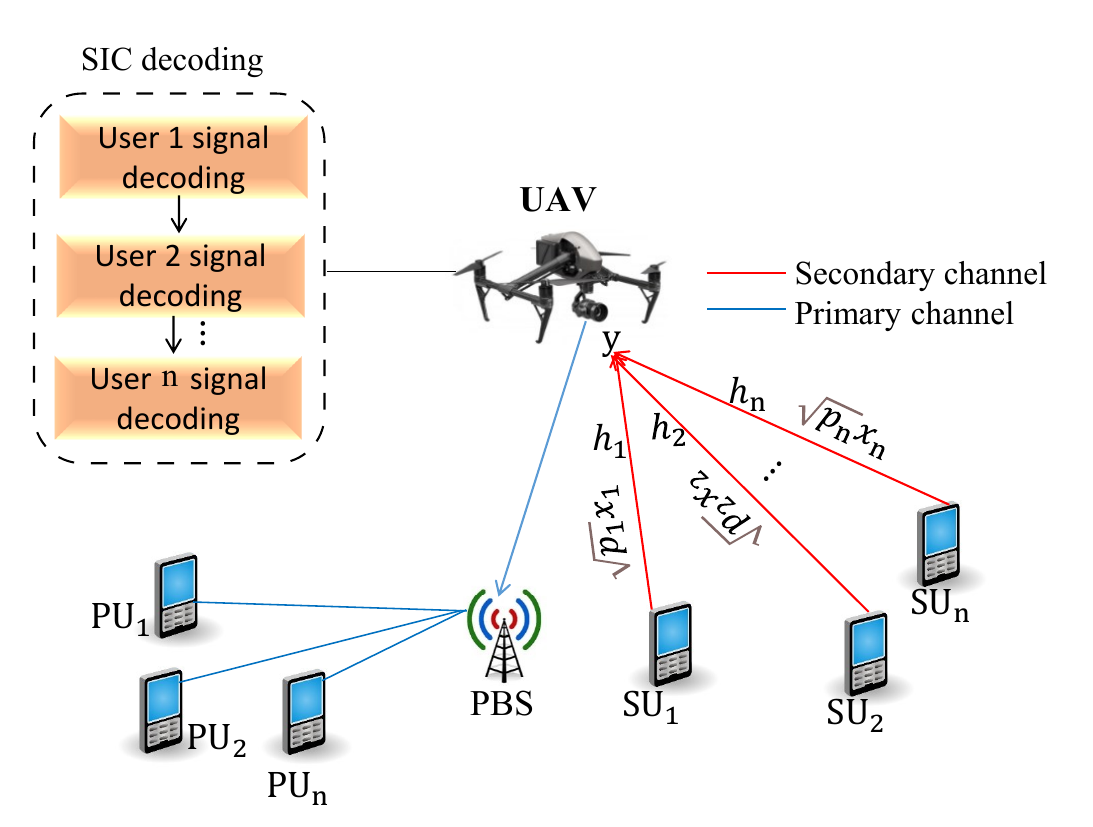}
    \caption{System model with uplink NOMA signaling.}
    \label{fig_system_model}
\end{figure}

As illustrated in Fig.~\ref{fig_system_model}, we examine a multi-channel uplink Cognitive-NOMA system that encompasses a primary network (PN) and a secondary network (SN), with a UAV positioned centrally and hovering above randomly moving secondary users (SUs). In practice, a single and hovering UAV could be used to provide communication services to emergency responders in the event of a disaster. This could be used to coordinate the response, communicate with victims, or provide medical assistance. The PN consists of a primary base station (PBS) that serves primary users (PUs) over the primary channels in a time-slotted manner. The SN consists of a UAV that assists the PBS and serves a set of SUs.
Let $\mathcal{N}$ denote the set of SUs and $\mathcal{K}$ represent the number of sub-channels in the network, expressed as $\mathcal{N}=\lbrace1,2,\cdots,N\rbrace$ and $\mathcal{K}=\lbrace1,2,\cdots,K\rbrace$, respectively. We assume non-interference among the different sub-channels due to the orthogonality provided by frequency division.

In the uplink, each SU $n$ transmits its signal to the UAV on subchannel $k$ with assigned transmit power $p^{k}_{n}$  and channel gain $g^{k}_{n}$. By using QPSK modulation, the system can maintain a certain level of performance and minimize the impact of interference compared to higher-order modulation schemes. This is particularly important in scenarios with multiple NOMA users, where the signals from different users may interfere with each other.
Let $\mathcal{U}_{k}\triangleq \{{n \in \mathcal{N}:p^{k}_{n} > 0}\}$ denote the set of SUs that are multiplexed on sub-channel $k$ and $|\mathcal{U}_{k}|$ represents the cardinality of that set. In each transmission time slot, the channel of a specific SU remains constant but changes independently in each period or episode. The UAV, equipped with Active-GDBN, can continuously update its policy online based on new observations of the channel state information (CSI). This adaptability allows the UAV to handle variations in the wireless channel conditions and adjust its actions accordingly. 
For simplicity, we assume a line-of-sight (LOS) channel and adopted the free-space path loss (FSPL) model as defined by \cite{wu2018joint}. Thus, the distance $d_{u,n}$ from the UAV $u$ to ground SUs $n$ at a given time instance $t$ is expressed as:
$d_{u,n} = \sqrt{h^{2} + \lVert\mathbf{q}_u(t)-{\mathbf{w}_n}\rVert^2}$,
where $h$ is the UAV's altitude, the horizontal coordinate of the UAV is represented by $\mathbf{q}_u(t)$, and $\mathbf{w}_n$=$[{x}_n \ {y}_n]^T$ denotes the horizontal coordinate for the mobile ground $n$-th SU. 
Similarly, the link power gain from UAV to SUs is given by:

\begin{equation} \label{channel_model_UAV_SU}
    h_{n}(t) = g_{n}^{k}(t) \Omega_{n}(t),
\end{equation}
where $g_{n}^{k}(t)$ is the large-scale power gain, accounting for path losses and shadowing, and is calculated as follows:
\begin{equation}
    g_{n}^{k}(t)=\rho_0{d^{-2}_{n,u}(t)=\frac{\rho_0}{{h^2}+\lVert\mathbf{q}_u(t)-{\mathbf{w}_n}\rVert^2}},
\end{equation}
where the link power gain at a reference distance is $\rho_0$. In \eqref{channel_model_UAV_SU}, $\Omega_{n}(t)$ is the small-scale fading coefficient, which follows a Rician distribution with a non-central chi-square probability density function (PDF) \cite{azari2016optimal}.
To ensure a minimum distance $\Delta_{y}$ between the superimposed SUs' signal, each SU is mapped to a unique QPSK constellation at the transmitter. This minimum distance is well-spaced to minimize interference and ensure successful SIC decoding at the receiver.
By using the learned generative model, the UAV can make informed predictions about each user's signal and estimate their contributions to the observed optimal superimposed signal. 
The UAV will perform SIC by actively adapting its actions online to decode $x_1$ first, which is the SU with the strongest channel gain, subtract it from the total received signal $\mathrm{y}_{t,k}$, and treat the other signals ($x_2$ to $x_{M}$) as interference. The UAV performs subsequent SIC and the next user with a stronger channel gain is decoded. The uplink SIC is performed in decreasing order of channel gain.
The achievable data rate $\mathrm{R}_{k,n}$ in the uplink is expressed as:
\begin{equation}
   \mathrm{R}_{k,n} \triangleq b_{k} \log_2 \bigg(1+\frac{p^{k}_{n}g^{k}_{n}}{\sum^{\vert\mathcal{U}_{k}\vert}_{j=\sigma^{-1}_{k}(n)+1}{p^{k}_{\sigma_{k}(j)}}{g^{k}_{\sigma_{k}(j)}}+{\eta_{n}^{k}}}\bigg).
\end{equation}
Our objective is to ensure a maximum sum-rate subject to power constraints while assuring the maximum number of allowable SUs per sub-channel. The maximization problem can be formulated mathematically as follows:
%
\begin{subequations} \label{eq_opt_sumRate}
\begin{align}
\max_{{p}_{n}^{k}} \quad & \sum^{K}_{k=1} \sum_{n=1}^{N}{}{\mathrm{R}_{k,n}} \label{eq_opt_sumRateA}\\
\textrm{s.t.} \quad & {\sum^{\vert\mathcal{U}_{k}\vert}_{k=1}} {p^k_n} \leq {p}_{max}, \quad {n\in \mathcal{N}}, {k \in \mathcal{K}} \label{eq_opt_sumRateB}\\
  & {p^k_n} \geq 0, \quad {n \in \mathcal{N}},{k \in \mathcal{K}} \label{eq_opt_sumRateC}\\
  & \vert\mathcal{U}_{k}\vert \leq M, \quad k \in \mathcal{K} \label{eq_opt_sumRateD}\\
  & p^{k}_{n} \leq {p}^{k,n}_{max}, \quad {n \in \mathcal{N}},{k \in \mathcal{K}}. \label{eq_opt_sumRateE}
\end{align}
\end{subequations}
Constraint \eqref{eq_opt_sumRateB} defines the maximum allowed total power budget for each SU, which cannot surpass $p_{max}$. \eqref{eq_opt_sumRateC} specifies that the power allocation for each SU on each sub-channel is non-negative. \eqref{eq_opt_sumRateD} restricts the maximum number of SUs multiplexed on a particular sub-channel to $M$. \eqref{eq_opt_sumRateE} sets power restrictions for each sub-channel.
The optimization task in \eqref{eq_opt_sumRateA} is nonconvex, and solving the global optimal using heuristic approaches is computationally infeasible. Therefore, we propose an active inference-based approach that efficiently learns the optimal subchannel and power allocation policy.

\section{Proposed method for joint sub-channel and power allocation} \label{Sec_proposedMethod}
We describe active inference as a partially observable Markov decision process (POMDP) \cite{friston2015active}. In a given time instance $t$, the actual state of an environment $\mathrm{\tilde{S}}_{t} \in \mathbb{R}^{d_s}$ changes according to a random transition process $\mathrm{\tilde{S}}_{t} \sim \mathrm{Pr}(\mathrm{\tilde{S}}_t \vert \mathrm{\tilde{S}}_{t-1},\boldsymbol{\mathcal{A}})$, where $\boldsymbol{\mathcal{A}}$ ${\in}$ $\mathbb{R}^{d_a}$ represents the actions of an agent (UAV).  The actual environmental state is usually hidden from the agent, but the agent can only infer them through observations $\mathrm{\tilde{Z}}_{t}$ ${\in}$ $\mathbb{R}^{d_z}$, given by $\mathrm{\tilde{Z}}_{t}\sim\mathrm{Pr}(\mathrm{\tilde{Z}}_{t}\vert\mathrm{\tilde{S}}_{t})$. As a result, the agent works with beliefs about the hidden state $\mathrm{\tilde{S}}_{t}$.

Under the active inference framework, the relationship between the UAV and its environment can be described as a 6-element tuple ($\boldsymbol{\mathrm{\tilde{S}_{t}}}$, $\boldsymbol{\mathrm{\tilde{X}_{t}}}$, $\boldsymbol{\mathcal{A}}$, $\boldsymbol{\mathrm{T}_{\tau}^{pu}}$, $\boldsymbol{\Pi_{\tau}^{a}}$, $\boldsymbol{\tilde{Z}_{t}}$), where $\boldsymbol{\mathrm{\tilde{S}_{t}}}$ and $\boldsymbol{\mathrm{\tilde{X}_{t}}}$ are sets of the environmental hidden states that include noise, PUs and/or SUs. $\boldsymbol{\mathcal{A}}=\{\mathcal{A}_{}^{[\mathrm{f}]}, \mathcal{A}_{}^{[p]}\}$ is the action space containing all the possible sub-channel decisions and initial power allocation values. 
$\boldsymbol{\mathrm{T}_{\tau}^{pu}}$ is the time-varying transition model for PUs. $\boldsymbol{\Pi_{\tau}^{a^{}}}$ is the Active Inference-table that encodes the state-action pair and $\boldsymbol{\tilde{\mathrm{Z}}_{t}}$ is the set of $K$ sensory signals. 
\subsubsection{Radio Environment Representation}
\begin{figure}[t!]
\centering
\begin{minipage}[b]{.493\linewidth}
\centering
    \includegraphics[height=3.0cm]{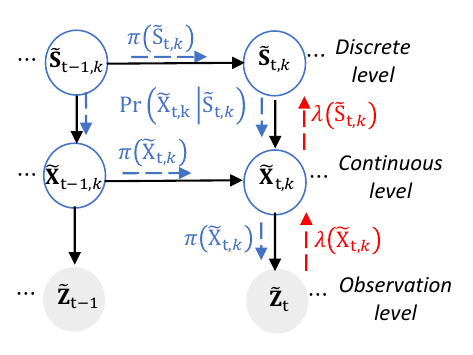}
\\[-1.0mm]
{\scriptsize (a)}
\end{minipage}
\begin{minipage}[b]{.493\linewidth}
    \centering
    \includegraphics[height=4.0cm]{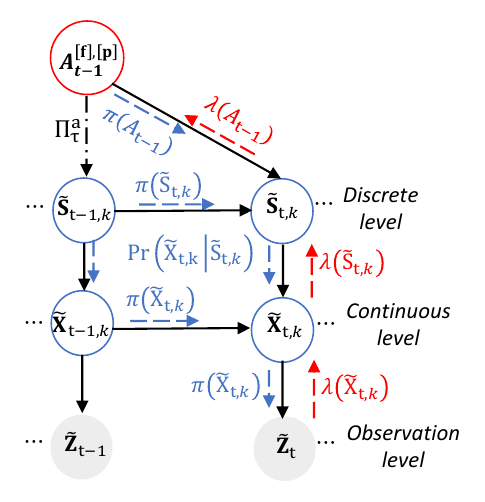}
    \\[-1.0mm]
    {\scriptsize (b)}
\end{minipage}

\caption{Graphical representations of the proposed method: (a) GDBN, (b) Active-GDBN: As depicted in sub-figure (b), the highest level of the hierarchy is the active states ($A_{t-1}^{[\mathrm{f}],[\mathrm{p}]}$), which indicate the joint actions of the UAV. Representing the joint sub-channel and power allocation variables using Active-GDBN enables us to describe the dynamic causal structure of the radio environment at discrete and continuous states, facilitated by constant message passing and belief updating. The blue arrows denote prior messages, while the red arrows represent future messages. In essence, the past and future states are constantly represented over time as new evidence becomes available. 
The joint actions ( $A_{t-1}^{[\mathrm{f}]}$ for discrete sub-channel selection) and ( $A_{t-1}^{[\mathrm{p}]}$ for continuous power allocation) affect the present states $\mathrm{\tilde{S}}_{t,k}$, $\mathrm{\tilde{X}}_{t,k}$ at time $t$ on sub-channel $k$ and defines the present observation $\mathrm{\tilde{Z}}_{t}$ which depends on the previous states $\mathrm{\tilde{S}}_{t-1,k}$, $\mathrm{\tilde{X}}_{t-1,k}$.}
    \label{fig_activeDBN}
\end{figure}

The UAV can observe $K$ sensory signals expressed as:
$\boldsymbol{\tilde{\mathrm{Z}}_{t}} {=} \{\mathrm{\Tilde{Z}}_{t,1}, \mathrm{\Tilde{Z}}_{t,2}, \dots, \mathrm{\Tilde{Z}}_{t,K}\}$, which correspond to $K$ sub-channels. 
In addition, we describe the radio environment using a generalized hierarchical state-space model, which includes the following components:
\begin{equation} \label{eq_discrete_model}
    \mathrm{\tilde{S}}_{t,k}^{(e)} = \textrm{f}(\mathrm{\tilde{S}}_{t-1,k}^{(e)}) + \mathrm{w}_{t,k},
\end{equation}
\begin{equation} \label{eq_continuous_model}
    \mathrm{\Tilde{X}}_{t,k}^{(e)} = \mathrm{C}\mathrm{\tilde{X}}_{t-1,k}^{(e)}+\mathrm{D}\mathrm{U}_{\mathrm{\tilde{S}}_{t,k}^{(e)}}+\mathrm{w}_{t,k},
\end{equation}
\begin{equation} \label{eq_observation_model}
    \mathrm{\Tilde{Z}}_{t,k} = \mathrm{H} \big(   \mathrm{\Tilde{X}}_{t,k}^{(1)}+\dots+\mathrm{\Tilde{X}}_{t,k}^{(M)}+\mathrm{\Tilde{X}}_{t,k}^{(pu)}\big)+\mathrm{v}_{t,k}.
\end{equation}
In \eqref{eq_discrete_model}, the discrete random variables describing the discrete sate clusters of the physical signal, the sub-channel carrying the signal and its power level are denoted by $\scriptsize \mathrm{\tilde{S}}_{t,k}^{(e)}$. 
Also, $\textrm{f}(.)$ is a non-linear function that expresses how $\mathrm{\tilde{S}}_{t,k}^{(e)}$ evolve over time as a function of $\mathrm{\tilde{S}}_{t-1,k}^{(e)}$ and $\mathrm{w_{t,k}}$ is the process noise, such that $\mathrm{w}_{t,k} {\sim} \mathcal{N}(0, \Sigma_{\mathrm{w}_{t,k}})$. The dynamic equation defined in \eqref{eq_continuous_model} expresses how the Generalized States (GS) $\mathrm{\Tilde{X}}_{t,k}^{(e)}$ evolve over time as a function of $\mathrm{\Tilde{X}}_{t-1,k}^{(e)}$ and $\mathrm{\tilde{S}}_{t,k}^{(e)}$ where $e\in\{no,pu,c\}$, $no$, $pu$, and $c$ stands for noise, PU and the $M$ superimposed signals, respectively. $\mathrm{C}$ and $\mathrm{D}$ represent the dynamic and control matrices, respectively, and $\mathrm{U}_{\mathrm{\tilde{S}}_{t,k}^{(e)}}$ is the control vector.
The observation model in
\eqref{eq_observation_model} describes dependence of the sensory signals on the hidden GS.
The hierarchical dynamic models formulated in terms of stochastic processes in \eqref{eq_discrete_model}, \eqref{eq_continuous_model}, and \eqref{eq_observation_model} are structured in a graphical GDBN as depicted in Fig.~\ref{fig_activeDBN}. 
The procedure includes an offline phase (i.e., the UAV's perception of desired observation), and the UAV is equipped with a hierarchical GDBN at discrete and continuous states to learn a generative model of the network, as depicted in Fig.~\ref{fig_activeDBN}(a). 
Due to the Markov separation between the UAV and the external world, the UAV learns an optimal policy of discrete sub-channels and continuous power by taking into account network conditions and user position.
Fig.~\ref{fig_activeDBN}(b) denotes the online active inference phase, where the UAV performs joint actions (i.e., dynamically selects continuous power $A_{t-1}^{[\mathrm{p}]}$ and discrete sub-channels $A_{t-1}^{[\mathrm{f}]}$) to reach the desired observation.
\subsubsection{Perceptual Learning of Preferred Observations}
At the beginning of the learning stage, the UAV is equipped with an initial model similar to the Unmotivated Kalman Filter (UKF) which assumes that the environmental states evolve along with static rules and relies on \eqref{eq_continuous_model} to predict the continuous environmental states where $\mathrm{U}_{\mathrm{\tilde{S}}_{t,k}^{(e)}}{=}0$ \cite{9829873}. 
The UAV’s memory produces initial errors known as generalized errors (GEs) \cite{9858012}. 
The GEs are further used to learn new models incrementally. We used the Growing Neural Gas (GNG) unsupervised clustering algorithm to learn the GDBN model that receives the GEs and generates discrete state clusters. 
Similarly, the time-varying transition matrix $\Pi_{k,\tau}$ is learned by estimating the transition probability $\mathrm{Pr}(\mathrm{\tilde{S}}_{t,k}^{(e)}{\rvert}{\mathrm{\tilde{S}}_{t-1,k}^{(e)}}, \tau)$.
The UAV repeats the previous learning procedure to learn distinct vocabularies that represent the various entities, such as, noise, PU, SU, and the combined signals generated from multiple SUs. 
\subsubsection{Active Inference Phase}
The UAV's decision-making depends on the state-action pair encoded in $\boldsymbol{\Pi_{\tau}^{[\mathrm{f}]^{}}}$, a time-varying matrix encoding the probabilistic dependencies between states and discrete actions, and $\boldsymbol{\Pi_{\tau}^{[\mathrm{p}]^{}}}$, a time-varying matrix encoding the probabilistic dependencies between states and continuous actions.
\paragraph{Action selection process} Initially, during the first iteration, the UAV performs random sampling to select the discrete actions as every possible discrete action has the same probability ($\frac{1}{K}$) of being chosen and selects the initial continuous action $A_{t-1}^{[\mathrm{p}]}=A_{0}^{[p]}$ for power allocation. The selected actions indicate what will be the next discrete and continuous environmental states $\mathrm{\tilde{S}}_{t,k}$, $\mathrm{\tilde{X}}_{t,k}$ which are characterized by $\mathrm{Pr}(\mathrm{\tilde{S}}_{t, {k}}^{}|\mathrm{\tilde{S}}_{t-1,{k}}^{}, A^{[\mathrm{f}]}_{t-1})$ and $\mathrm{Pr}(\mathrm{\tilde{X}}_{t, {k}}^{}|\mathrm{\tilde{X}}_{t-1,{k}}^{}, A^{[p]}_{t-1})$.
In the successive iterations, the UAV can adjust the action selection process by predicting implicitly the future activity of PUs according to $\boldsymbol{\mathrm{T}_{\tau}^{pu}}$ and skipping the resources that are expected with high probability to be occupied by PUs. 
By utilizing a modified Markov Jump Particle Filter (M-MJPF)\cite{9322583}, the UAV is capable of predicting the outcomes of its actions. The M-MJPF utilizes a switching model, employing Particle Filtering (PF) for prediction and updating in the discrete state, and Kalman Filtering (KF) for prediction and updating in the continuous state. Through dynamic causal relationships, a top-down inference can be distinguished from a bottom-up inference. Additionally, the UAV observes and senses the unselected sub-channels to determine their state (occupied or vacant) and detect the activity of primary users (PUs) in the spectrum. This information is used to enhance future decision-making processes.
Time-based inter-slice top-down predictive messages $\pi(\mathrm{\tilde{X}}_{t,k})$ and $\pi(\mathrm{\tilde{S}}_{t,k})$ is based on the information acquired in the dynamic model. The intra-slice bottom-up inference is built on the likelihood function and consists of backward propagated messages $\lambda(\mathrm{\tilde{X}}_{t,k})$ and $\lambda(\mathrm{\tilde{S}}_{t,k})$ towards the discrete level. The prediction at the continuous level depends on the discrete level. For each particle propagated at the discrete level, a KF is activated to predict the equivalent continuous level $\mathrm{\tilde{X}}_{t,k}$. 
PF propagates $L$ particles equally weighted based on the proposal density encoded in transition matrix $\Pi_{k}$.
After receiving the new observation, diagnostic messages propagate in a bottom-up manner to update the belief in hidden variables at the different hierarchical levels (continuous and discrete states).
\paragraph{Abnormality measurements and action evaluation} 
The continuous level abnormality indicator calculates the similarity between the two messages entering the node $\mathrm{\tilde{X}}_{t,{k}}^{}$, namely, $\pi(\mathrm{\tilde{X}}_{t,{k}}^{})$ and $\lambda(\mathrm{\tilde{X}}_{t,{k}}^{})$
to understand how much the observation supports the predictions, according to:
\begin{equation}
\scriptsize
    \boldsymbol{\Upsilon_{\mathrm{\tilde{X}}_{t,{k}}^{}}} = -\ln \bigg( \mathcal{BC}\big(\pi(\mathrm{\tilde{X}}_{t,{k}}^{}),\lambda(\mathrm{\tilde{X}}_{t,{k}}^{})\big) \bigg)= \int \sqrt{\pi(\mathrm{\tilde{X}}_{t,{k}}^{})\lambda(\mathrm{\tilde{X}}_{t,{k}}^{})} d\mathrm{\tilde{X}}_{t,{k}}^{},
    \label{CLA_abn_reference_model}
\end{equation}
where $\mathcal{BC}$ is the Bhattacharyya coefficient.
The UAV can decide whether the allocated actions were good or wrong by comparing the multiple abnormalities. 
\paragraph{Updating of action selection process}
The UAV receives sensory signals to perceive the radio environment and modifies the environment through its actions. It then infers the effects of the executed actions, both discrete and continuous, through observations. By selecting appropriate actions, the UAV can adapt its strategy and determine its future behavior by minimizing the GEs given by:
\begin{equation} \label{eq_error_discrete_action}
    \tilde{\mathcal{E}}_{A_{t-1}} = \big[{A_{t-1}}, \dot{\mathcal{E}}_{A_{t-1}} \big] = \big[ {A_{t-1}}, \lambda(A_{t-1}) - \boldsymbol{\pi}(A_{t-1}) \big].
\end{equation}
\section{Simulation Results and Discussion} \label{Sec_simulationResults}
In this section, we assess the performance of our proposed Active-GDBN. We fix the radius of the cell $R$ to 1000 meters. Within the cell, there is a PBS, and a UAV is positioned in the middle, serving $N$ SUs. We assumed that three PUs are actively occupying three sub-channels and the other sub-channels are vacant. Table I summarizes the network parameters. 

\begin{figure}[ht!]
    \centering
    \hspace*{-0.5cm}
    \includegraphics[height=2.9cm]{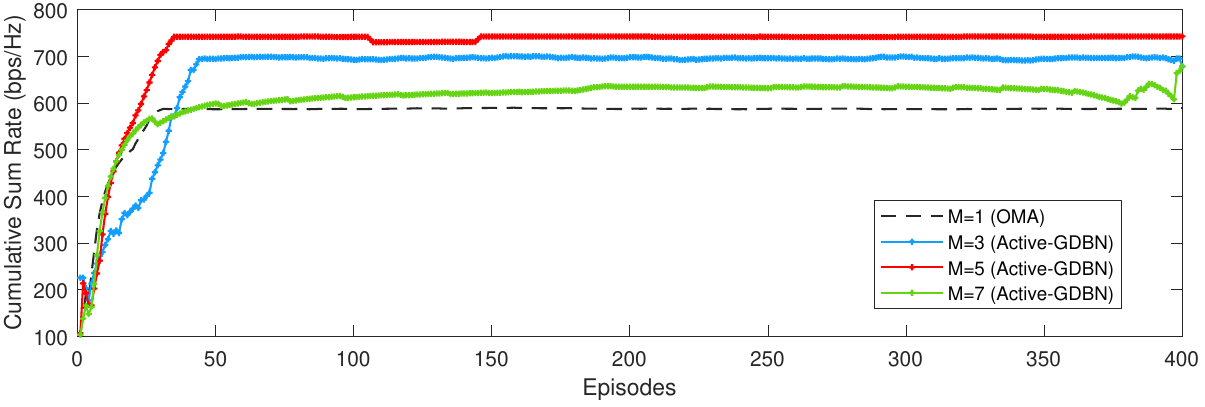}
    \caption{Convergence of Active-GDBN with different numbers of multiplexed SUs when $M$ = 5, $P_{th} = 1$, $P_{max} = 20$ Watts.}
    \label{fig_sumRate_comparison}
\end{figure}

\begin{table}[!ht]
\begin{center}
\caption{Simulation Parameters}
\scalebox{0.9}{%
\begin{tabular}{ |l||l| }
 \hline
 Cell radius & 1000 m \\ 
 \hline
 Min. distance from UAV to SUs & 100 m \\ 
 \hline
 Modulation scheme of PUs & BPSK  \\ 
 \hline
 Modulation scheme of SUs & QPSK  \\ 
 \hline
 Path loss model & Free-space-path-loss \cite{wu2018joint}  \\ 
 \hline
 Noise power & $-174$ dBm/Hz  \\ 
 \hline
 System Bandwidth, $B_{w}$ & 1.4 MHz  \\ 
 \hline
 Number of sub-channels  $K$ & 6  \\ 
 \hline
 Number of SUs $N$ & $20$  \\ 
 \hline
 Power budget of SUs $P_{max}$ &  20 W\cite{wang2018optimal} \\ 
\hline
 Number of SUs multiplexed per sub-channel $M$ & $M = [1,3,5,7]$,  \\ 
 \hline
 Power difference threshold $P_{th}$ & $1$  \\ 
 \hline
 Learning rate of GNG clustering & $0.01$  \\ 
 \hline
\end{tabular}}
\end{center}
\end{table}
Fig.~\ref{fig_sumRate_comparison} illustrates the convergence performance of the proposed algorithm where the sum rate values are plotted versus the number of episodes for different numbers of SUs ($M$) multiplexed per sub-channel.  When $M = 1$, the problem reduces to orthogonal multiple access (OMA). As revealed, it takes within zero and fifty episodes to converge for all possible values of $M$. Moreover, the sum rate increases with an increasing number of SUs. The proposed algorithm achieved a maximum number of $5$ SUs multiplexed per sub-channel. Also, as we increased the number of SUs to $7$, the proposed algorithm shows performance degradation. 
This is because, beyond this limit, the symbols of the superimposed SU signals begin to overlap, making accurate SIC and signal decoding impossible for the UAV. Thus, the difference between the superimposed constellation points $\Delta_{y}$ is kept at a reasonable distance to avoid inter-symbol interference.
\begin{figure}[bt!]
    \centering
    \includegraphics[width=9.0cm]{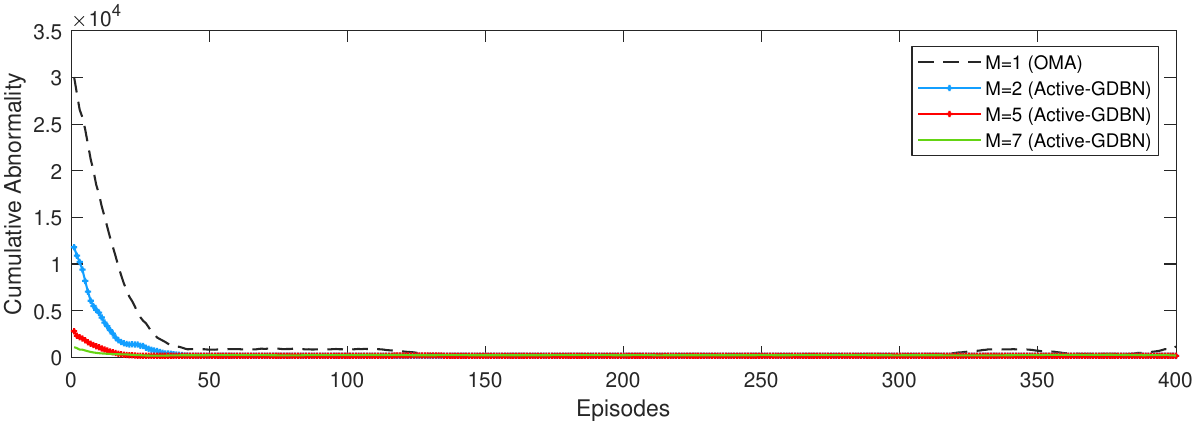}
    \caption{Cumulative Abnormality of the proposed Active-GDBN with different numbers of multiplexed SUs when $M$ = 5, $P_{th} = 1$, $P_{max} = 20$ Watts.}
    \label{fig_CumAbn}
\end{figure}

Fig.~\ref{fig_CumAbn} shows the cumulative abnormality results of the proposed algorithm, validating Fig.~\ref{fig_sumRate_comparison}. We transform the sum rate maximization problem into abnormality minimization. As a result, Active-GDBN abnormality minimization is equivalent to Active-GDBN reward (i.e., cumulative sum rate) maximization.

\begin{figure}[t!]
    \centering
    \includegraphics[width=9.0cm]{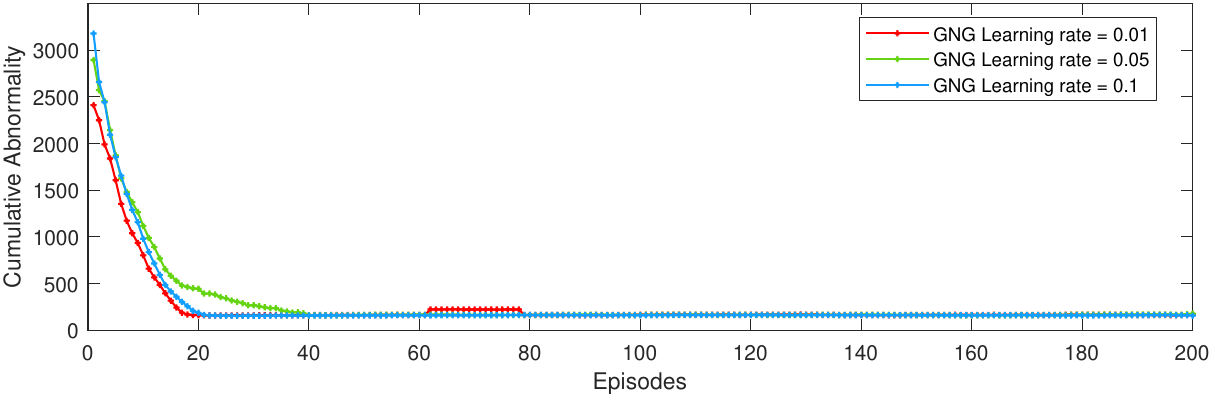}
    \caption{Cumulative abnormality of the proposed Active-GDBN with different GNG learning rates when $M = 5$, $P_{th} = 1$, $P_{max} = 20$ Watts.}
    \label{fig_variable_Lrate}
\end{figure}

Fig.~\ref{fig_variable_Lrate} reveals the cumulative abnormality value as a function of different GNG learning rates during offline training. As is evident, setting the learning rate to 0.01 results in the fewest episodes required to attain the minimum cumulative abnormality. Therefore, to achieve faster convergence, we set the learning rate to 0.01 for all simulation settings.

\begin{figure}[b!]
    \centering
    \includegraphics[width=9.0cm]{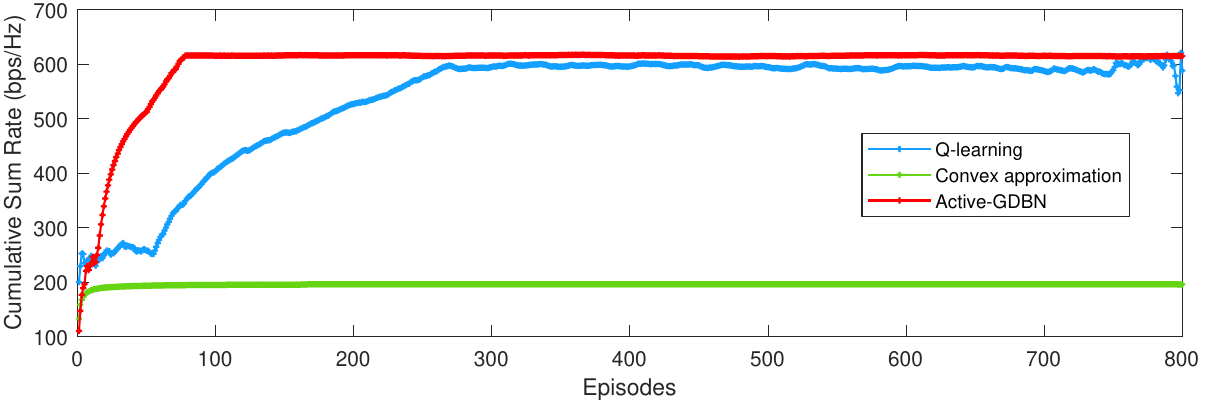}
    \caption{Cumulative sum rate comparison of the proposed Active-GDBN with benchmark schemes when $M = 5$, $P_{th} = 1$, $P_{max} = 20$ Watts.}
    \label{fig_GEs_SUs}
\end{figure}
To compare the performance of the proposed Active-GDBN, we adopt and modify the Q-learning algorithm from \cite{huang2020online} and the successive convex approximation technique from \cite{guo2020weighted}. As clearly indicated in Fig.~\ref{fig_GEs_SUs}, the proposed method surpasses the Q-learning scheme to reach a better and more stable sum rate in fewer episodes. This is because, in each time step, the UAV detects abnormalities, implying a mismatch between the preferred observations and the predictions due to the performed actions. As a result, the UAV exploits the errors in each episode to learn how to take better actions that minimize future abnormalities. Moreover, the proposed Active-GDBN performs dynamic continuous power allocation to SUs. Due to the strong influence of negative rewards on Q-learning, it requires more training episodes to achieve a significant improvement in the sum rate. The low sum rate performance of convex approximation is due to its inability to learn from experience and adapt its strategies to the time-varying radio environment.

\begin{figure}[ht!]
    \centering
    \includegraphics[width=7.0cm]{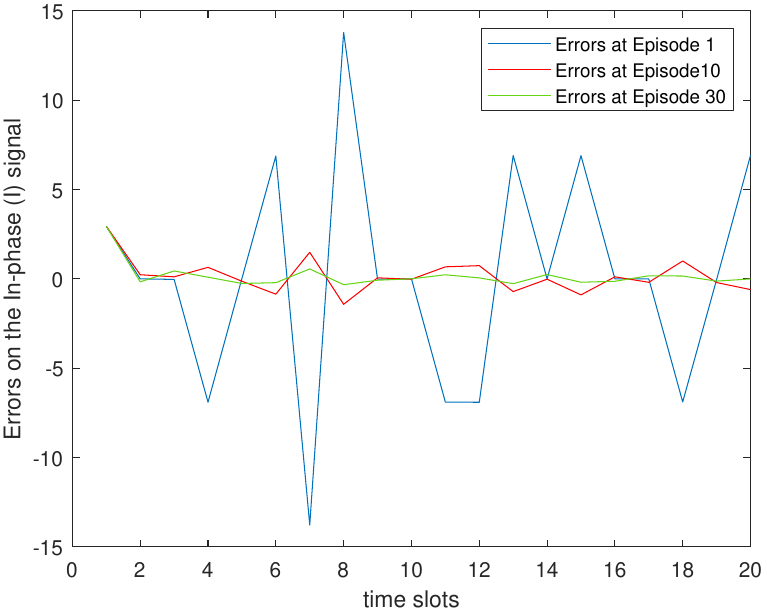}
    \caption{An example of the In-phase prediction errors with varying training episodes per time slot when $M = 5$, $P_{th} = 1$, $P_{max} = 20$ Watts.}
    \label{fig_Errors_Inphase}
\end{figure}
 Fig.~\ref{fig_Errors_Inphase} indicates an example of the errors on the in-phase component of the predicted combined SUs’ signal. The initial errors are high in the first episode (blue line) because the UAV’s initial beliefs about the SUs’ positions and channel conditions are not accurate. The red line shows that the errors have decreased significantly after 10 episodes, as the UAV has learned to adapt its resource allocation policies and belief updating. The errors decrease to a minimum after 30 episodes (green), as the UAV continues to learn and improve by exploiting the generalized prediction errors.
\section{Conclusion} \label{Sec_conclusion}
In this study, we investigate the joint sub-channel and power allocation problem in an uplink UAV-assisted cognitive NOMA network. We propose an active inference-based algorithm, called Active-GDBN, to solve the sum rate maximization problem. Due to network dynamics and practical limitations on the number of users multiplexed per sub-channel, the problem is usually difficult to solve analytically. As a result, we use a generalized state space model to characterize the dynamic network environment and transform the problem into an abnormality minimization problem. After performing extensive simulations, the results reveal the effectiveness of our proposed algorithm over the benchmark schemes in terms of cumulative sum rate. Future research will examine the effect of the UAV's trajectory in relation to the mobility of SUs and high-order modulation schemes.
 
\bibliographystyle{IEEEtran}
\bibliography{References}

\vspace{12pt}

\end{document}